\documentclass[RNAAS]{aastex631}

\begin{document}

\title{The Effect of Photometric Errors on the Measured Width of the Main Sequence in Star Clusters}

\correspondingauthor{Steven Spangler}
\email{steven-spangler@uiowa.edu}

\author[0000-0002-4909-9684]{Steven R. Spangler}
\affiliation{Department of Physics and Astronomy, University of Iowa}

\begin{abstract}
This paper deals with the effect of errors in the B and V magnitudes, or measurements in any other color system, on the width of the main sequence in a color-magnitude (Hertzsprung-Russell) diagram. The width is defined as the dispersion in apparent (or absolute) magnitude at a fixed, measured photometric color. I find that the dispersion is larger than might be thought {\em a priori}. A statistical analysis is presented which demonstrates that the error in the magnitude residual from a linear approximation to the main sequence is Gaussian, but with a standard deviation which is much larger, in general, than the errors in the individual B and V magnitudes. This result is confirmed by a Monte Carlo simulation of a main sequence population with specified errors in B and V magnitudes, and can be explained on the basis of simple algebraic arguments.  
\end{abstract}

\section{1. Introduction} 
This is a second paper dealing with the width of the main sequence (MS) in an open star cluster, or any sample of main sequence stars.  By the width, I mean the range of apparent or absolute magnitudes measured for the same value of the color, surface temperature, or stellar mass.  The first paper, \cite{Spangler25a} \citep[a detailed description is given in][]{Spangler25b} considered the effect of undetected binary stars in this phenomenon.  The current paper is concerned with the more pedestrian issue of the effect of errors in the magnitudes in two colors such as B and V. 
  
To provide a context for this paper, it is the second in a planned set of three, with the goal of measuring or placing limits on long-term luminosity variations of solar-type stars.  As briefly mentioned in \cite{Spangler25b}, the idea is that low-amplitude intrinsic variations in the luminosity of solar-type stars might be  one of the mechanisms contributing to  the broadening of the main sequence of an open star cluster. My plan is to apply the results of \cite{Spangler25a} and the present paper to the case of the open star cluster M67 in the third paper in this series.

\section{2. Basic Concepts}
The analysis is carried out in a plane with Cartesian coordinates $(x,y)$. The connection to the color-magnitude diagram of an open star cluster is provided by defining these coordinates as 
\begin{eqnarray}
y \equiv m_V  \\ \nonumber
x \equiv (m_B - m_V)
\end{eqnarray}
where $m_B$ and $m_V$ are the apparent magnitudes in the B and V Johnson filters, respectively.  However, I again emphasize that the analysis given here could be applied to a color-magnitude diagram (CMD) with any set of filters, such as the Sloan filters or those used on the Gaia spacecraft. The analysis presented below would also apply to a Hertzsprung-Russell (HR) diagram in which absolute magnitudes are plotted versus color index, with photometric errors propagated into the absolute magnitude values.   

To simplify the analysis, I will approximate the main sequence (MS) by a linear function in the color-magnitude $(x,y)$ diagram.  In the case of negligible dispersion by any mechanism, the MS is then approximated by
\begin{equation}
y = ax + b
\end{equation}
with $a$ and $b$ being empirical fit parameters.  I adopt Equation (2) primarily for the substantial simplification it allows, but it also turns out to be a reasonable approximation to the main sequence in the vicinity of solar-type stars.  

\section{3. Statistics of Variations}
In reality, the measured values of $x$ and $y$ will have a contribution due to measurement error.  From now on, I adopt shorthand variables $v \equiv m_V$, $b \equiv m_B$.  A graphical representation of the effect of errors is given in Figure 1. I consider a cell in the $(x,y)$ plane, centered at $(x,y)$ and with dimensions $(dx,dy)$.  I let the variables $(x',y')$  be the corresponding values for a star without photometric errors, i.e. a star on the noise-free main sequence.  

\begin{figure}[h]
\begin{center}
\includegraphics[scale=0.50,angle=0]{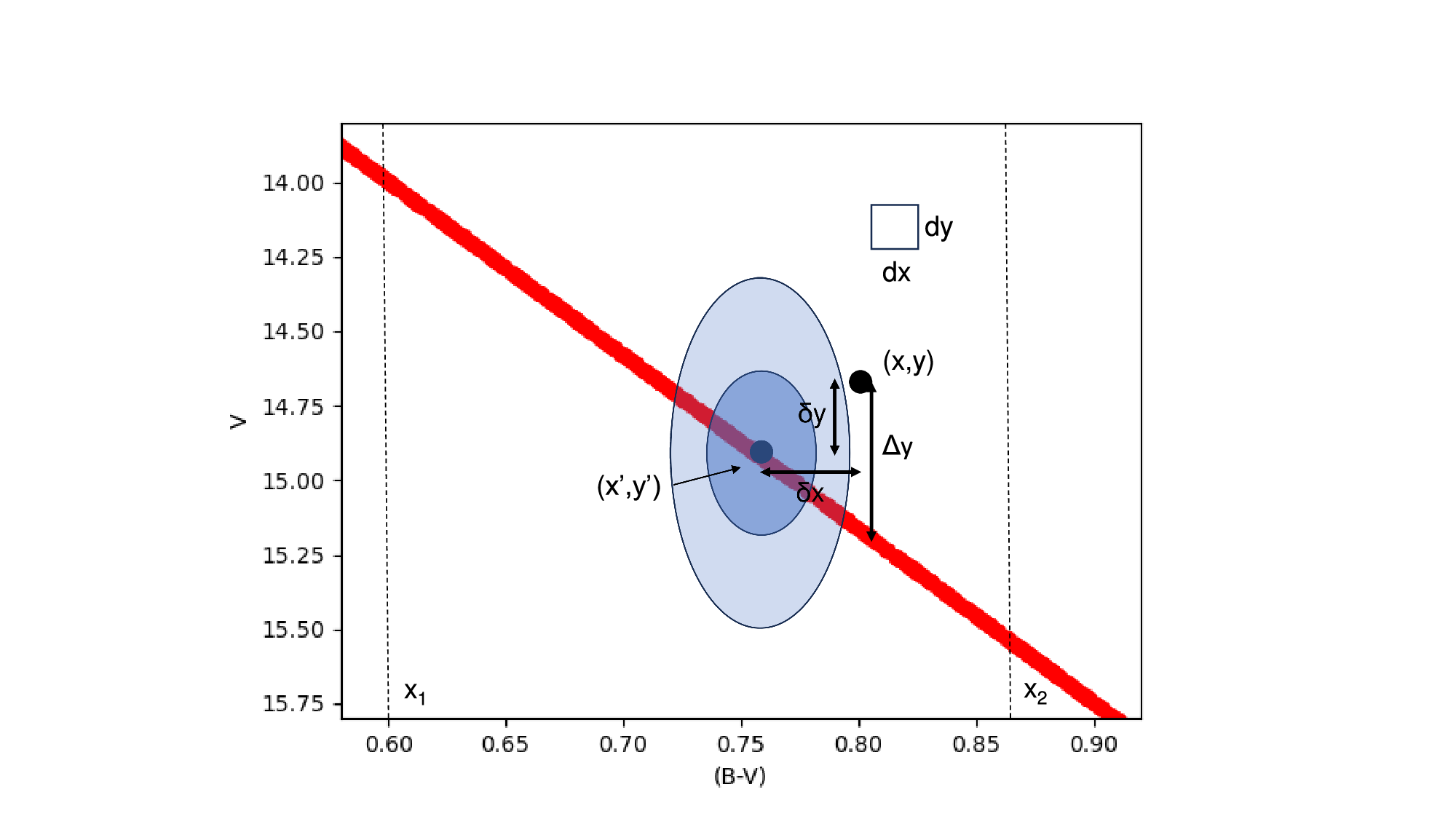}
\caption{An illustration of the probability of stars with true color and magnitude $(x',y')$ (see definition in Equation (1)) being observed with color and magnitude $(x,y)$. The shading represents the probability density $p(x,y,x',y')$ defined in Equation (3). All variables are defined in the text, e.g. $\Delta y$ is defined in Equation (32). The values of $x_1$ and $x_2$ shown here are for purposes of illustration, and do not necessarily correspond to the values chosen in the calculations.}
\end{center}
\end{figure}
The fundamental equation in my analysis is that for the differential number of stars in the aforementioned cell in $(x,y)$ space, $\Delta n(x,y,x',y')$, 
\begin{equation}
\Delta n(x,y,x',y') = \sigma(x',y') p(x,y,x',y') ds dxdy
\end{equation}
with $ds$ being an incremental distance along the line approximating the MS, and $\sigma(x',y')$ being the linear density of stars along the MS.  The probability density $p(x,y,x',y')$ expresses the probability that a star with coordinates $(x',y')$ will produce, via the errors in $v$ and $b$, a measured value of $(x,y)$.  As will be discussed now, the probability density $p(x,y,x',y')$ is not totally trivial because the variables $(x,y)$ are correlated. 

Given the shorthand variables introduced above, Equation (1) is reexpressed as $y=v$, $x = b -v$.  It is reasonable to model $(b,v)$ as independent random variables, but this is not the case for $(x,y)$.  Expressing the independent, random variations in $(b,v)$ by $(\delta b, \delta v)$, the corresponding random variations in $(x,y)$ are 
\begin{eqnarray}
\delta y \equiv y-y'= \delta v \\ \nonumber
 \delta x \equiv x-x' = \delta b - \delta v 
\end{eqnarray} 
Finally, I adopt the reasonable approximation
\begin{equation}
p(x,y,x',y') = p_{xy}(\delta x,\delta y)
\end{equation}
which should be valid, at least for a limited portion of the MS that contains solar-type stars.

I now consider the transition from $p_{bv}(\delta b,\delta v) \rightarrow p_{xy}(\delta x,\delta y)$. As noted above, it is reasonable to assume $(\delta b,\delta v)$ as independent random variables, for which I assume a dual Gaussian model, 
\begin{equation}
p_{bv}(\delta b,\delta v) = \frac{1}{2 \pi \sigma_B \sigma_V} \exp \left[ - \left( \frac{(\delta b)^2}{2 \sigma_B^2} + \frac{(\delta v)^2}{2 \sigma_V^2} \right) \right]
\end{equation}

The approach employed is to carry out a coordinate conversion $(\delta b,\delta v)$ to two other variables$(\delta x(\delta b,\delta v),\delta y(\delta b,\delta v))$.  The formula relating the corresponding probability density functions is derived in  \cite{Beckmann67}, Section 2.4, and utilizes the conservation of probability, 
\begin{equation}
p_{xy}(\delta x,\delta y)d (\delta x) d (\delta y) = p_{bv}(\delta b, \delta v) d (\delta b) d (\delta v)
\end{equation}
This leads to \cite[][Section 2.4]{Beckmann67}
\begin{equation}
p_{xy}(\delta x,\delta y) = p_{bv}(\delta b, \delta v) \left| \frac{d(\delta b) d(\delta v)}{d(\delta x) d(\delta y)} \right|
\end{equation}
where the Jacobian of the transformation is given by 
\begin{equation}
 \left| \frac{d(\delta b) d(\delta v)}{d(\delta x) d(\delta y)} \right|\equiv  \left| \left| \begin{array}{cr}
\frac{db}{dx} & \frac{db}{dy} \\ \frac{dv}{dx} & \frac{dv}{dy}   \end{array} \right| \right|
\end{equation}
that is, the absolute magnitude of the determinant of the $2 \times 2 $ matrix given  within the outer set of brackets on the right hand side of Equation (9).  In Equation (9) I have used the shorthand notation $\frac{db}{dx} \equiv \frac{d (\delta b)}{d (\delta x)}$, etc.  

Use of Equation (4) shows that $ \left| \frac{d(\delta b) d(\delta v)}{d(\delta x) d(\delta y)} \right| =1$, so I have
\begin{equation}
p_{xy}(\delta x,\delta y) = p_{bv}(\delta b(\delta x,\delta y), \delta v(\delta x,\delta y))
\end{equation}
Using Equation (6), I have
\begin{equation}
p_{xy}(\delta x,\delta y) = \frac{1}{2 \pi \sigma_B \sigma_V} \exp \left[ - \left( \frac{(\delta x + \delta y)^2}{2 \sigma_B^2} + \frac{(\delta y)^2}{2 \sigma_V^2} \right) \right]
\end{equation}
Equation (11) illustrates that, although $p_{bv}(\delta b,\delta v)$ is separable (into a product of two independent terms), $p_{xy}(\delta x,\delta y)$ is not.  

Continuing with algebraic ``productive playtime'', Equation (11) becomes
\begin{equation}
p_{xy}(\delta x,\delta y) = \frac{1}{2 \pi \sigma_B \sigma_V} \exp \left[ - \left( \frac{(\delta x)^2}{2 \sigma_B^2} + \left[ \frac{1}{2 \sigma_B^2} +  \frac{1}{2 \sigma_V^2} \right](\delta y)^2 +\frac{2 (\delta x) (\delta y)}{2 \sigma_B^2} \right) \right]
\end{equation}
The first two terms in the argument of the exponential in Equation (12) represent a pdf in which $x$ and $y$ would be independent.  The third term represents the effect of the correlation of $x$ and $y$, or equivalently, $m_V$ and $(B-V)$.  
\section{4. The Probability Density in the $(x,y)$ Plane}
To obtain the probability in the $(x,y)$ plane, I substitute Equation (12) into Equation (3), and perform the integral over $ds$.  Since $(x',y')$ lie on a straight line, there is a simple relation for $ds$,
\begin{eqnarray}
ds = \sqrt{(dx')^2 + (dy')^2} \\ \nonumber
 = \sqrt{1 + \left(\frac{dy'}{dx'} \right)^2}dx' = \sqrt{1 + a^2}dx'
\end{eqnarray}
given that the main sequence is being approximated by a linear function.  

In what follows, I will also assume that the linear density of stars along the main sequence is a constant, $\sigma(x',y') = \sigma_0$, and further note that $\sigma_0 L = N_{\ast}$, where $N_{\ast}$ is the number of stars on the main sequence, or more specifically, the number in the part of the main sequence that is of interest.  In the case of the present research program, this means main sequence stars that are also solar-type stars. 

The somewhat artificial parameter $L$ can be eliminated in favor of the range of color (and therefore stellar mass) of interest, $L = \sqrt{1 + a^2}(x_2 - x_1)$.  Here $x_1, x_2$ are the smallest and largest values of color in the region of interest on the main sequence, and are important parameters in the analysis.  

Substituting all of this into Equation (3) yields the following expression for the number of stars in a cell in the $(x,y)$ plane
\begin{equation}
\frac{d^2n}{dxdy}=\frac{N_{\ast}}{(x_2 - x_1)}\left( \frac{1}{2 \pi \sigma_B \sigma_V} \right)\int^{x_2}_{x_1}dx' \exp [-\alpha(x,y,x')]
\end{equation}
where $\delta x$ and $\delta y$ have been converted back to $\delta x = x-x'$, $\delta y = y-y' = y - y_0(x')$, with $y_0(x') = ax' + b$. To be absolutely clear in the notation of Equation (14), $   \frac{d^2n}{dx dy} \equiv \frac{\Delta n}{dx dy}$. In Equation (14) the integration over $ds$ has been converted to an integral over $x'$.  Reference to Equation (12) shows that the argument of the exponential in the integrand, $\alpha(x,y,x')$ is given by
\begin{equation}
2 \sigma_B^2 \alpha(x,y,x') = (x-x')^2 + \left[1 + \frac{\sigma_B^2}{\sigma_V^2} \right](y-y_0(x'))^2 + 2(x-x')(y-y_0(x'))
\end{equation}
For brevity of notation, I henceforth define $R$ by 
\begin{equation}
R^2 \equiv 1 + \frac{\sigma_B^2}{\sigma_V^2}
\end{equation}

After soul-deadening but completely straightforward algebraic manipulation, Equation (15) can be converted into a form quadratic in the variable of integration, $x'$.  
\begin{equation}
2 \sigma_B^2 \alpha(x,y,x') = A(x')^2 - 2B(x') + C + 2(x-x')(y-ax'-b)
\end{equation}
Equation (17) is written in this form to isolate the last term, which arises from the correlation of the variables $(x,y)$.  The quadratic terms absent the last term would be the correct expression for $\alpha(x,y,x')$ in the case of uncorrelated $(x,y)$.  The coefficients $B$, $C$ are functions of $(x,y)$, $A$ is a constant (expressions given in Equation (19) below).  

The last (fourth) term in Equation (17) can also be expanded in powers of $x'$, and added to the appropriate quadratic terms of Equation (17), leading to the overall quadratic expression, 
\begin{equation}
2 \sigma_B^2 \alpha(x,y,x') = A'(x')^2 - 2B'(x') + C' 
\end{equation}
with 
\begin{eqnarray}
A'=A+2a=1+R^2a^2+2a \\ \nonumber
B'=B+ax+(y-b)=x+R^2a(y-b)+ax+(y-b) \\ \nonumber
C'=C+2(y-b)x=x^2+R^2(y-b)^2+2(y-b)x
\end{eqnarray} 
where again, $B'=B'(x,y)$, $C'=C'(x,y)$.

A final manipulation is to divide both sides of Equation (18) by $A'$, a constant, yielding 
\begin{equation}
\frac{2 \sigma_B^2 \alpha(x,y,x')}{A'} = (x')^2 - 2\beta'(x') + c' 
\end{equation}
where $\beta' \equiv \frac{B'}{A'}$, $c' \equiv \frac{C'}{A'}$.

The goal now is to substitute Equation (20) into Equation (14), perform the integration over $x'$ (thus adding up the contributions from all points along the noiseless main sequence), and obtain the stellar density in the $(x,y)$ plane.  

To evaluate the integral, I complete the square in Equation (20), i.e. let it be rewritten as 
\begin{equation}
\frac{2 \sigma_B^2 \alpha(x,y,x')}{A'} = (x')^2 - 2\beta'(x') + c' +d' -d' 
\end{equation}
, such that 
\begin{equation}
(x'-\beta')^2 = (x')^2 - 2\beta'(x') + c' +d' 
\end{equation}
Straightforward algebra shows this is satisfied if 
\begin{equation}
d'=(\beta')^2-c' 
\end{equation}
with $d'=d'(x,y)$.

With this substitution, Equation (14) becomes 
\begin{equation}
\frac{d^2n}{dxdy}=\frac{N_{\ast}}{(x_2 - x_1)}\left( \frac{1}{2 \pi \sigma_B \sigma_V} \right) \exp \left[ \frac{A'd'}{2 \sigma_B^2} \right] \int^{x_2}_{x_1}dx' \exp [-\frac{A'(x'-\beta')^2}{2 \sigma_B^2}]
\end{equation}

\subsection{4.1 Evaluation of Integral Over Variable $x'$}
I now consider the integral in Equation (24), noted by $I$
\begin{equation}
I (x,y) \equiv \int^{x_2}_{x_1}dx' \exp [-\frac{A'(x'-\beta')^2}{2 \sigma_B^2}]
\end{equation}
The obvious choice in evaluating this integral is to adopt a change of variables, 
\begin{eqnarray}
x' \rightarrow \xi \equiv \sqrt{\frac{A'}{2}} \left( \frac{1}{\sigma_B}\right)(x'-\beta')  \\ \nonumber
dx'= \sqrt{\frac{2}{A'}} \sigma_B d \xi
\end{eqnarray}
Substitution of Equation (26) into (25) gives
\begin{eqnarray}
I = \sqrt{\frac{2}{A'}} \sigma_B \left( \frac{\sqrt{\pi}}{2} \right) \left[ \frac{2}{\sqrt{\pi}} \int_{\xi_1}^{\xi_2} d \xi e^{-\xi^2} \right] \\ \nonumber
=  \sqrt{\frac{2}{A'}} \sigma_B \left( \frac{\sqrt{\pi}}{2} \right) \left[ \mbox{erf}(\xi_2) - \mbox{erf}(\xi_1) \right]
\end{eqnarray}
where $\mbox{erf}(\xi_2)$ is the Error Function of argument $\xi_2$.
\subsection{4.2 Considerations Regarding the Arguments of the Error Functions}
Although Equation (27) for the integral $I$ is in closed form, it is potentially a complicated function of $(x,y)$ through the arguments of the error function.  

To evaluate Equation (27), consider first the upper argument of the error function, 
\begin{equation}
\xi_2 = \sqrt{\frac{A'}{2}} \left( \frac{1}{\sigma_B} \right) (x_2 - \beta')
\end{equation} 
As seen in Figure 1, $x_2$ is the upper extremum of color index, by definition larger than all other points along the main sequence that is being considered.  Using the definition of $\beta'$ in Equation (19) and (20), and assuming that all values of $y$ lie close to the main sequence, i.e. $y \sim ax +b$, it turns out that $\beta' \simeq x$, so that 
\begin{equation}
\xi_2 \simeq \sqrt{\frac{1+R^2a^2 +2a}{2}} \left( \frac{x_2 - x}{\sigma_B} \right) 
\end{equation}
For the main sequence, the slope $a \gg 1$, and by definition, $R \geq 1$.  As a result, the term in the square root of Equation (29) will be a number $\gg 1$.  Furthermore, for almost all points (except at the very end of the interval $x_1 \leq x \leq x_2$, $x_2 - x \gg \sigma_B$, i.e. the photometry errors are small compared with the entire color range considered, so the second term in expression (29) will also be much larger than unity.  

As a result, except for values of $x'$ well within a single photometric error of $x_2$, $\xi_2 \gg 1$, and as a result, $\mbox{erf}(\xi_2) \simeq 1$ to a very good approximation.  

The same reasoning applied to $\xi_1$ leads to the conclusion $\xi_1 \ll -1$, and $\mbox{erf}(\xi_1) \simeq -1$.  The net result is that the density in the color magnitude diagram, Equation (24), becomes 
\begin{equation}
\frac{d^2n}{dxdy} = \frac{N_{\ast}}{(x_2 - x_1)} \left( \frac{1}{\sigma_V} \right) \frac{1}{\sqrt{2 \pi A'}} \exp \left[ \frac{A'd'}{2 \sigma_B^2} \right] 
\end{equation}
\section{5. Final Expression for Probability Density Function in Color-Magnitude Plane}
To obtain a clear and useful result from Equation (30), I need an identity for the variable $d'$, defined in Equation (23). Start with $\beta'$, which is defined following Equation (20). Use of the definitions of $B'$ and $A'$ (Equation (19), leads to 
\begin{equation}
\beta' = \frac{1}{A'}\left[ (1+a)x + (1+R^2a)(y-b) \right]
\end{equation}
In this paper, I am not interested in $y$ per se (the apparent or absolute magnitude of star in the CMD), but rather the departure above or below the nominal, noise-free main sequence.  Let $\Delta y$ be defined via the relation
\begin{equation}
y=(ax+b)+\Delta y
\end{equation}
with $\Delta y$ now being the difference between the observed magnitude of a star with (observed) color index $x$ and the calculated magnitude of a star with the same color index on the nominal main sequence (see Figure 1 for an illustration). The variable $\Delta y$ defined in Equation (32) and used henceforth differs from the similar variable written down in Equation (4). 

Substitution of Equation (32) into Equation (31) yields 
\begin{equation}
\beta' = x+\left[ \frac{1+R^2a}{A'} \right](\Delta y)
\end{equation}
Following the same steps, the parameter/function $c'$ is
\begin{equation}
c' = \frac{1}{A'}(x^2 + 2(y-b)x + R^2(y-b)^2)
\end{equation}
The same expression Equation (32) can be used to convert Equation (34) to a more useful form.  

Algebraic manipulation of the expression for $d'$ in Equation (23), utilizing the relations in Equations (33) and (34) leads to the remarkably simple result
\begin{equation}
d'=-\frac{1}{(A')^2} [R^2-1](\Delta y)^2
\end{equation}
The definition of $R^2$ in Equation (16) insures that $R^2 > 1$, and $d'< 0$, as required for a normalized probability density function.  

Substitution of Equation (35) into Equation (30) gives
\begin{equation}
\frac{d^2n}{dxdy} = \frac{N_{\ast}}{(x_2 - x_1)} \left( \frac{1}{\sigma_V} \right) \frac{1}{\sqrt{2 \pi A'}} \exp \left[ -\frac{(R^2-1)(\Delta y)^2}{2 A' \sigma_B^2} \right] 
\end{equation}
Remarkably, the expression has no dependence on $x$, meaning the distribution of magnitudes above and below the nominal main sequence is independent of color index. It should be recalled that at the outset of this analysis, I assumed that the linear density of stars along the noise-free main sequence is a constant, i.e. $\sigma(x',y')=\sigma_0$.  
\section{6. The $\Delta y$ Probability Density Function} 
To complete the derivation, three steps are undertaken. (1) A width parameter 
\begin{equation}
\tilde{\sigma} \equiv \sqrt{\frac{A'}{R^2-1}}\sigma_B
\end{equation}
is defined. 
(2) Equation (36) is integrated over $x$ from $x_1 \rightarrow x_2$, yielding a pdf for the brightness residual $p(\Delta y)$. (3) The total number of stars $N_{\ast}$ is divided out, yielding a properly normalized pdf.  The result is
\begin{equation}
p(\Delta y) = \frac{1}{\sqrt{2 \pi} \tilde{\sigma}} \exp \left[-\frac{(\Delta y)^2}{2 \tilde{\sigma}^2}  \right]
\end{equation}
Equations (37) and (38) present the main result of this paper.  Given the approximation that a portion of the main sequence of a star cluster, or other assembly of stars can be approximated by a line on a color-magnitude diagram, the probability density function of variations $\Delta y = \Delta m$ or $\Delta M$ above or below the main sequence (at a fixed color index $x$) is a Gaussian with standard deviation $\tilde{\sigma}$.  

To appreciate the significance of Equation (37) for the standard deviation, refer to Equation (16) for $R^2$, and Equation (19) for $A'$. The important result from Equation (37) is that the effective standard deviation of the magnitude deviations, $\tilde{\sigma}$, is much larger than the photometric error $\sigma_V$ or $\sigma_B$.  The dispersion in apparent or absolute magnitude at a fixed color is related to a typical photometric error ($\sigma_B$) by the factor $\sqrt{\frac{A'}{R^2-1}}$. The denominator inside the square root will typically be a number of order unity.  However, the numerator is large.  

It may be noticed from Equation (19) that the quantity $A'$, and therefore $\tilde{\sigma}$ is dependent on $a$, which may be positive or negative for a general linear relationship.  This then indicates that $\tilde{\sigma}$ would be different for hypothetical, noise-free linear relationships with $|a|$ the same, but with different signs of $a$.  This result is a consequence of the correlation between the random variables $x$ and $y$ introduced by the defining relations Equation (1).  The above statements were verified by simulations using the same computer program described in the following section. For the main sequence in a color-magnitude diagram (the physical application of interest), $a > 0$.

A qualitative explanation for the result contained in Equations (37) and (38) is simple. For the main sequence part of the HR diagram, there is a large change in $M_V$ for a relatively small change in $(B-V)$ (or equivalent color), a consequence of the steepness of the mass-luminosity relation.  On a standard plot with linear axes, the main sequence is a line with a steep slope.  As a consequence, a small error in the color leads to a large error in the expected magnitude. 
\section{7. Simulation of the Broadening of the Observed Main Sequence by Photometric Errors}
The analysis to this point contains a number of assumptions, approximations, and mathematical deductions, any of which could be the source of a blunder which would cause Equations (37) and (38) to be in error.  To test the accuracy and fidelity of these equations, I undertook a numerical simulation of the process of photometric errors superposed to an infinitely narrow main sequence.  

The simulation was carried out in a plane with Cartesian coordinates $(x,y)$.  The $y$ coordinate is taken to be the V band apparent magnitude $m_V$, and $x \equiv m_B - m_V$, where $m_B$ is the B band apparent magnitude, as was the case in my analytic treatment of the problem.  

The simulation was carried out via the following steps.  
\begin{enumerate}
\item A linear model for the noise-free main sequence was chosen according to Equation (2). I chose $a=5.83$ and $b=10.50$.  This pair was used because it is a reasonable approximation to the main sequence part of the HR diagram for M67 (to be discussed in the planned third paper in this series).  However, for the present application, I am only seeking values which are approximately appropriate for the main sequence in the vicinity of the Sun and solar-type stars.  The slope ($a$) is obviously the more important of the two parameters.  
\item A number $N_s$ values of $x'$  were chosen uniformly on the interval $x_{lo} \leq x' \leq x_{hi}$.  These were chosen using the Python subroutine numpy.random.uniform.  The integer $N_s$ is taken to be the number of simulated stars, and the interval $(x_{lo},x_{lo})$ is taken to be slightly larger than the color range $x = (B-V)$ of interest.  
\item Values for $y$ corresponding to the randomly selected $x$ values from step \# 1 were assigned a value using Equation (2), and the chosen values for $(a,b)$. This corresponding set of $N_s$ values of $(x,y)$ represent the noise-free MS, and may be considered the variables $(x',y')$ in Section 3 and Figure 1. 
\item Two arrays of dimension $N_s$ were created, $e_s$ and $e_t$, and populated with independent, Gaussian-distributed random numbers with standard deviation $\sigma_s$ and $\sigma_t$, respectively, and zero mean. The random variables $e_s$ and $e_t$ were calculated using the Python subroutine numpy.random.normal.  The random variable $e_t$ models the photometric measurement error in the V band apparent magnitude, and $e_s$ is the corresponding error in the B apparent magnitude.  
\item The simulation proxies for the measured magnitude and color were calculated as 
\begin{eqnarray}
x_{obs} = x + (e_s - e_t)  \\  \nonumber
y_{obs} = y + e_t
\end{eqnarray}
\item The resultant simulated data set displays a ``clipped'' distribution of values for $x$ near the edges of the simulation, $x \simeq x_{lo}$ or $x \simeq x_{hi}$.  This is due to the fact that contributions are missing from ``stars'' outside the range $(x_{lo},x_{lo})$.  For this reason, I chose a subsample of the simulated stars with $x_1 \leq x \leq x_2$, where the range $(x_1,x_2)$ corresponds to the portion of the MS of interest (shown in Figure 1).  In this project, that range is of solar-type stars. The number of stars in the clipped distribution is noted by $N_{s2}$. The resultant set of $(x_{obs},y_{obs})$  values corresponds to the simulated open cluster CMD diagram, including the effect of errors in the B and V magnitudes.  
\end{enumerate}
This simulation was undertaken to verify the correctness of Equations (37) and (38), but was also chosen to represent the open cluster M67. The parameters used in the simulation are given in Table 1.  \\
 
{\bf Table 1. Input Parameters to Simulation} \\ 
\begin{tabular}{||l|l|l||} \hline \hline
Model Parameter & Value & Definition-Commment \\ \hline
$N_s$ & 850 & number simulated stars \\ \hline
$a$ & 5.83 & slope of MS \\ \hline
$b$ & 10.50 & intercept of MS \\ \hline
$\sigma_s = \sigma_B$ & 0.015 & rms error, B \\ \hline
$\sigma_t = \sigma_V$ & 0.015 & rms error, V \\ \hline
$x_{lo}$ & 0.50 & lower limit, simulated color \\ \hline
$x_{hi}$ & 1.00 & upper limit, simulated color \\ \hline
$x_1$ & 0.60 & lower limit, analysis region \\ \hline
$x_2$ & 0.90 & upper limit, analysis region \\ \hline
$N_{s2}$ & 525 & number stars, analysis region \\ \hline \hline
\end{tabular} \\

The photoelectric errors used in this simulation are comparable to those cited by \cite{Montgomery93}, who presented B and V magnitude measurements for M67, and who analysed the HR diagram for that cluster (see Sections 2 and 3, and Figure 1 of \cite{Montgomery93}).  The \cite{Montgomery93} data were also used by \cite{Geller15} in their study of M67. 

The simulated HR diagram is shown in Figure 2.  This display shows, at least qualitatively, that the scatter of the simulated stars (blue dots) about the model MS (solid red line) at a specified color is substantially larger than the individual errors in $m_V$ and $m_V$.
\begin{figure}[h]
\begin{center}
\includegraphics[scale=0.50,angle=0]{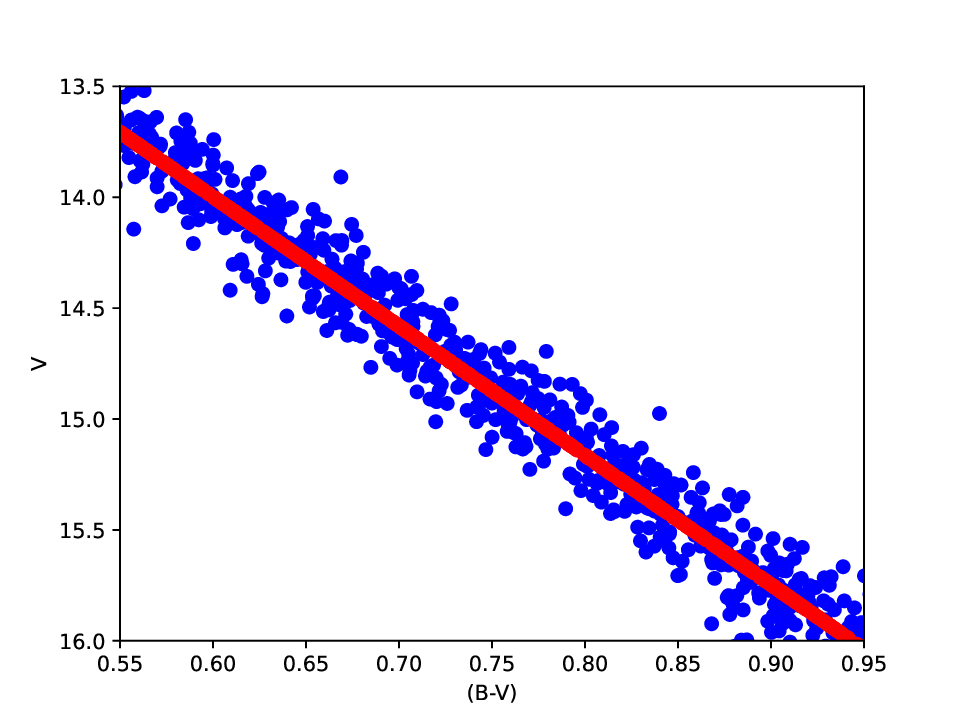}
\caption{Simulation of the main sequence with photometric errors in V and B. Parameters of the simulation are given in Table 1.}
\end{center}
\end{figure}

Given the results of this simulation, it is possible to test the validity of Equations (37) and (38).  Given the simulated data $(x_{obs},y_{obs})$, I calculated the quantity 
\begin{equation}
\Delta y = y_{obs} - (a x_{obs}+b)
\end{equation}
for all $N_{s2}$ stars in the clipped sample, as described in point \# 6 above.  
This corresponds exactly to the difference between the observed apparent magnitude of a star and the apparent magnitude that would be inferred on the basis of its observed color, as discussed in Sections 2-6.  
\begin{figure}[h]
\begin{center}
\includegraphics[scale=0.50,angle=0]{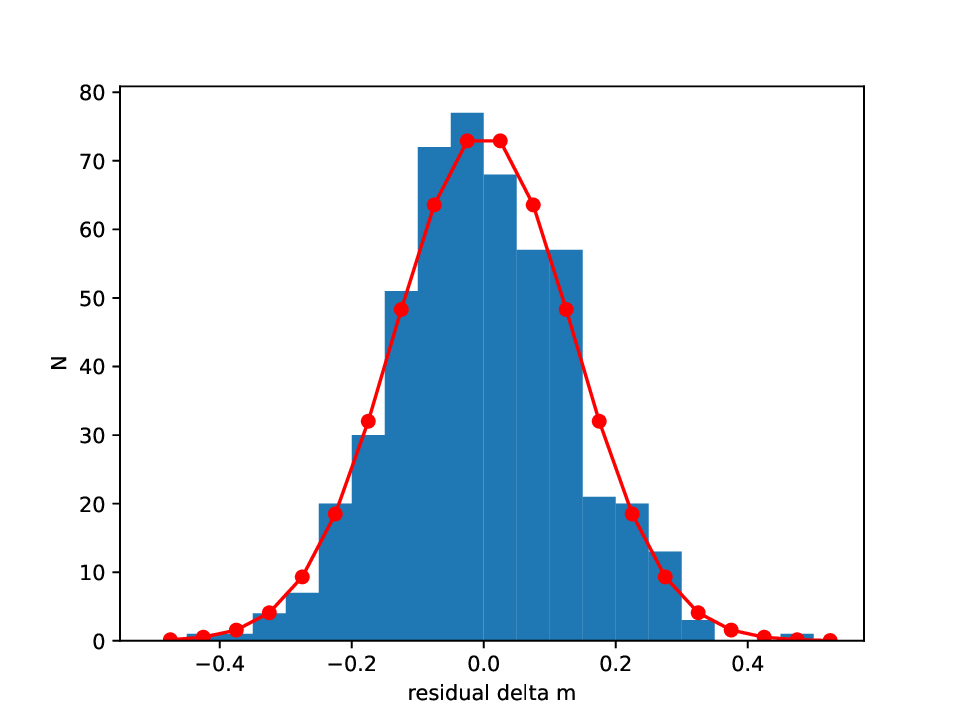}
\caption{The simulated pdf of the residual 
 magnitude $\Delta m$ with superposed Gaussian errors corresponding to expected noise. The red curve represents the analytic expression, Equation (38).}
\end{center}
\end{figure}
The results for this simulation are shown in Figure 3.  The blue histogram gives the distribution from the simulation.  The red points with connected lines represent the mathematical model described by Equations (37) and (38). Equation (38) has been converted to expected number of stars per bin, given the total number of stars in the simulated sample and the bin size.  I consider the agreement between the analytic pdf of Equation (38) and the results of the simulation to be satisfactory.  

A further test of the validity of Equation (38) was another simulation with different value of the slope $a$ and noise ratio $R$, resulting in a quite different value of the parameter $\frac{A'}{R^2-1}$.  The agreement between the Equation (38) and that simulation was equally good.  
\section{8. Implications of Results}
The main result of this paper is contained in Equations (37) and (38), with Equation (37) giving the rms width of the main sequence of a star cluster, due solely to photometric errors in the B and V magnitudes.  The same conclusion would result if other filter systems, like the Sloan filters, were used. The coefficient $\sqrt{\frac{A'}{R^2-1}}$ is considerable larger than unity for the main sequence, and in the normal case of similar errors in the different colors.  This is due to the steepness of the main sequence on a color-magnitude diagram.  

Given this steepness, my result can be understood on the basis of elementary algebra.  For a linear relation $y_0(x)$ with a large slope, a small error in the abscissa produces a large discrepancy from the expected value of the ordinate. This result is a fundamental consequence of the mass-luminosity relation for main sequence stars.  The color is a measured quantity which is a proxy for the stellar mass.  A relatively small error in the color therefore leads to a substantial error in the estimated luminosity or absolute magnitude, resulting in a star being inferred to be much brighter or fainter than expectation.  

A conclusion of this paper is that, if intrinsic luminosity variations of solar-type main sequence stars at an interesting level are to be retrieved from measurement of the width of the main sequence, measurements with very small photometric errors (i.e $\sigma_B \ll 0.010$ magnitude) will be required. Fortunately, space astronomy missions of the last one to two decades, with precision photometry of tens of thousands or millions of stars, have made this feasible. 
\section{9. Summary and Conclusions}
\begin{enumerate}
\item I have obtained an expression for the width of the main sequence in a star cluster, defined here as the dispersion in apparent (or absolute) magnitude at a fixed color.  Although the notation assumes that the observations are made in the Johnson B and V system, the formulas are applicable to any set of astronomical filters. 
\item The analysis assumes that the noise-free main sequence can be approximated, over a color range of interest, by a linear relationship with slope $a$ and intercept $b$. 
\item Assuming that the errors in the magnitudes $m_V$, $m_B$ etc, are Gaussian-distributed with standard deviations $\sigma_V$ and $\sigma_B$, the dispersion above and below the main sequence, denoted by $\Delta y$, is also Gaussian-distributed with a standard deviation $\tilde{\sigma}=\sqrt{\frac{A'}{R^2-1}}\sigma_B$ (Equation 37), with the parameters $A'$ and $R$ defined in Equations (19) and (16), respectively. For the main sequence, the constant $\sqrt{\frac{A'}{R^2-1}} \gg 1$, meaning that the main sequence width is considerably larger than might be thought on the basis of the photometric errors. 
\item At an astrophysical level, this result is a consequence of the fact that for main sequence stars, color is a proxy for the stellar mass, and luminosity is a strong function of mass.  A small measurement error in color results in a relatively large error in predicted stellar magnitude.  
\item Given existing photometric data from ground-based telescopes, intrinsic variations in stellar luminosity must be several percent to perhaps 10 percent to be detectable.  
\item Placing truly interesting limits on long term luminosity variations of solar-type stars, consisting of variations at the level of a percent to a few percent, will require photometric measurements with $\sigma_V, \sigma_B \ll 0.010$ magnitudes. Such measurements should be available in archived data from existing spacecraft missions such as Gaia and Kepler.  
\end{enumerate}

\end{document}